\documentclass[twocolumn,showpacs,amsmath,amssymb]{revtex4-1}
\input{epsf}

\usepackage{graphicx}
\usepackage{dcolumn}
\usepackage{array}                      
\usepackage{bigstrut}
\usepackage{longtable}
\usepackage{rotating,booktabs}
\usepackage{booktabs,threeparttable}
\usepackage{color}
\usepackage{bm}
\def\bdot{\mbox{\boldmath{$\cdot$}}}
\usepackage{color}

\begin{document}

\title{Theory for the Rydberg states of helium: Results for $2 \le n \le 35$ and comparison with experiment for the singlet and triplet $P$-states}

\author{G. W. F. Drake$^{1,*}$}
\author{Aaron T. Bondy$^2$}
\author{Oliver P. Hallett$^1$}
\author{Benjamin C. Najem$^1$}

\affiliation { $^1$ Department of Physics, University of Windsor, Windsor, Ontario, Canada N9B 3P4}
\affiliation{ $^2$ Department of Physics and Astronomy, Drake University, Des Moines, Iowa 50301}

\begin{abstract}
High precision variational calculations in Hylleraas coordinates are presented for all singlet and triplet $P$-states of helium up to principal quantum number $n = 35$ with a uniform accuracy of 1 part in $10^{22}$ for the nonrelativistic energy.  Mass polarization, relativistic and quantum electrodynamic effects are included to achieve a final accuracy of $\pm$1 kHz or better for the ionization energy of the Rydberg states of $^4$He in the range $24\le n \le 35$.  The results are combined with 11 transition frequency measurements of Clausen et al.\ Phys.\ Rev. A  {\bf 111}, 012817 (2025) to obtain complementary measurements of the ionization energy of the $1s2s\;^3S_1$ state that do not depend on quantum defect extrapolations to the series limit.  The result from the triplet spectrum yields an ionization energy of  1152\,842\,742.728(6) MHz, which agrees with but is larger than the experimental value by 14 $\pm$17 kHz. However, it confirms a much larger 9$\sigma$ discrepancy of $0.468\pm0.055$ MHz with the theoretical ionization energy of Patk\'o\v{s} et al.\ Phys.\ Rev.\ A {\bf 103}, 042809 (2021). The results provide a test of the quantum defect extrapolation method at the level of $\pm$17 kHz. This revised version contains an additional table of spin-dependent matrix elements from the Breit interaction 
in the appendix for $n\le n \le 35$. The results are otherwise identical.
\end{abstract}

\date{\today}

\maketitle

Recent work has revealed a 9$\sigma$ disagreement between theory and experiment in the ionization energy of the metastable $1s2s\;^3S_1$ state of helium \cite{Clausen2021,Clausen2025}.  The origin of the discrepancy remains unexplained, even though helium is a fundamental three-body problem for which high precision calculations are available, including electron correlation, relativistic and quantum electrodynamic corrections for the $1s2s\;^3S_1$ state \cite{Yan1995,Pachucki2012,Pachucki2010,Zheng2017,Kato2018}. The experiments of Clausen et al.\ \cite{Clausen2025} involve direct measurements of the ionization energy of the $1s2s\;^3S_1$ state by extrapolation of the $1s2s\;^3S_1-1snp\;^3P_J$ $(J=0,1,2)$ transition frequencies to the series limit $n\rightarrow\infty$ by the use of quantum defect theory, beginning at $n=27$.  This experiment followed an earlier indirect measurement based on the $1s2s\;^1S_0-1snp\;^1P_1$ transition frequencies, beginning at $n=24$, together with the high precision measurement of the $1s2s\;^1S_0-1s2s\;^3S_1$ transition frequency \cite{Rengelink2018}.  The two measurements agree with each other to within their statistical uncertainty ($0.023 \pm 0.035$ MHz), and they confirm a discrepancy of $0.4774\pm0.053$ MHz, with experiment lying deeper than theory.  Measurements on the $^3$He isotope yield a parallel discrepancy of
$0.482 \pm0.053$ kHz \cite{Clausen2025c}.  The uncertainty of $\pm$0.053 MHz represents the theoretical uncertainty due to higher order QED contributions.      

The measurements of Clausen et al.\ rely on the use of quantum defect theory to extrapolate to the series limit in order to find the ionization energy of the $1s2s\;^3S_1$ state in question.  Although quantum defect theory is well established and widely used, it is not an exact fundamental theory \cite{DrakeAdv}.  An alternative would be to perform a direct comparison between theory and experiment for transition frequencies to the high-$n$ Rydberg states.  The comparison would be meaningful because the QED uncertainties decrease approximately in proportion to $1/n^3$, and so become negligibly small at $n = 24$ and beyond.  We showed in a previous publication \cite{Bondy2025} that it is indeed possible to perform high precision calculations for $n=24$ by the use of triple basis sets in Hylleraas coordinates.  This goes well beyond the previous limit of $n = 10$ \cite{Drake1992}, and represents a significant extension of the technology for high precision atomic theory into previously unexplored territory at the kHz level of accuracy (parts in $10^{12}$ for the total energy).  In the present paper, these same techniques are applied to obtain ionization energies of all singlet and triplet $P$-states up to $n=35$, including relativistic and QED corrections up to order $\alpha^3$ Ry ($\alpha^5mc^2$), and estimates of the $\alpha^4$ Ry term based on exact calculations of the dominant part due to singlet-triplet mixing.  The results enable direct determinations of the ionization energy of the $1s2s\;^3S_1$ state without the need for extrapolation to the series limit. 

The paper is organized as follows.  The following section describes the triple basis set method of constructing nonrelativistic wave functions in Hylleraas coordinates for the Rydberg $P$-states of helium.  Varational bounds are tabulated and compared with previous work.  In Sect. \ref{relativistic}
the formalism for relativistic and QED corrections is presented.  The dominant terms of $O(\alpha^4)$ and $O(\alpha^5)$ are calculated directly, and the remainder estimated by a $1/n^3$ scaling from the known $n=2$ case.  Section \ref{results} presents a tabulation of the calculated ionization energies for the $P$-states up to $n=35$. The ionization energy of the $1s2s\;^1S_1$ state is calculated for each of the 11 measured $1s2s\;^1S_0 - 1snp\;^1P_1$ or
$1s2s\;^3S_1-1snp\;^3P_{\rm c}$ transition frequencies, and compared with the results of a quantum defect extrapolation. For the sake of brevity in the spectroscopic notation, the one-electron labels are suppressed so that, for example, $2\,^1P_1$ denotes the state $1s2p\;^1P_1$. Units are atomic units throughout.  The value of the reduced mass Rydberg constant used to convert from atomic units to MHz is
\begin{equation}
R_M = (1-\mu/M) R_\infty
\end{equation}
with $R_\infty =  3289\,841\,960.250$ MHz, and $\mu/M =  1.370\,745\,6347\times10^{-4}$, where $\mu = Mm_{\rm e}/(M + m_{\rm e})$ is the reduced electron mass and $M$ is the bare alpha-particle mass.  

\section{Nonrelativistic Wave Functions}
\label{nonrelativistic}
The primary obstacle to extending high precision calculations to the high-lying Rydberg states of helium is the difficulty of obtaining sufficiently accurate wave functions for the complete three body problem of a nucleus plus two electrons.  Variational calculations typically suffer from a rapid decline in accuracy with increasing principal quantum number $n$, as is evident in the original ground-breaking calculations of Accad, Pekeris and Schiff \cite{Accad71} in Hylleraas coordinates.  The basis functions involve powers of $r_1$, $r_2$ and $r_{12} = |{\bf r}_1 - {\bf r}_2|$ for the radial coordinates of the two electrons.  Their result for the $5\,^1P$ barely exceeds the accuracy of a simple screened hydrogenic approximation $-2.02000$ E$_{\rm h}$.  Drake et al.\ \cite{Drake87,Drake88,DrakeMakowski88,DrakeYan92} showed that much improved accuracy could be obtained by ``doubling" the Hylleraas basis set so that it explicitly contains two variationally determined distance scales for the inner and outer electron. This served well for all states of helium up to $n=10$ and $L=7$ \cite{DrakeYan92} (including relativistic and QED corrections). However, to reach the level of $n=24$ and beyond, we have shown in our previous paper \cite{Bondy2025} that results of sufficient accuracy can be obtained by tripling the basis set so that each combination of powers is included three times with different exponential distance scales. In other calculations, Chi et al.\ \cite{Chi2025} have recently obtained results of encouraging accuracy up to $n = 27$ for the nonrelativistic energy by the use of of correlated B-spline basis functions with up to 64000 terms, although the accuracy begins to deteriorate beyond $n=20$.  The only other calculations to go beyond $n = 10$ are the ICI calculations of Nakashima et al.\ \cite{Nakashima2008} up to $n= 24$ for the nonelativistic energies of $S$-states.  In addition, very high precision calculations have been performed by Aznabaev et al.\ \cite{Aznabaev2018} for the nonrelativistic energies for all states up to $n = 4$, using stochastic all-exponential basis functions with up to 22000 terms of the form $\exp(-\alpha_ir_1-\beta_ir_2-\gamma_ir_{12})$.

 As discussed previously \cite{Bondy2025,Nistor2002,Petrimoulx},  the heliumlike Hamiltonian is diagonalized in a basis set constructed from basis functions of the form (for $P$-states)
 \begin{eqnarray}
 \phi_{i,j,k}(\alpha,\beta;{\bf r}_1,{\bf r}_2) &=& r_1^i\,r_2^{j+1}\,r_{12}^k\exp(-\alpha r_1 - \beta r_2)\cos\theta_2\nonumber\\
                                                &&\mbox{}\pm \mbox{ exchange}
 \end{eqnarray}
where ${\bf r}_1$ and ${\bf r}_2$ are the position vectors of electron 1 and electron 2 relative to the nucleus of charge $Z$, $\cos\theta_2$ gives the angular dependence on ${\bf r}_2$ appropriate to a $p$-electron, and $r_{12} = |{\bf r}_1-{\bf r}_2|$ is the inter-electron separation.  The ``exchange" term denotes an interchange of the labels 1 and 2, with the (+) sign for triplets and the ($-$) sign for singlets.  A variational wave function in a triple basis set then has the form
\begin{equation}
\psi({\bf r}_1,{\bf r}_2) = \sum_{p=1}^{3}\,\,\sum_{i,j,k}^{i+j+k \le \Omega_p} c_{i,j,k}^{(p)}\phi_{i,j,k}(\alpha_p,\beta_p;{\bf r}_1,{\bf r}_2)
\end{equation}
where the $c_{i,j,k}^{(p)}$ ar linear variational coefficients, and $\Omega_p$ controls the size of the basis set such that $i+j+k\le \Omega_p$ in each of the sectors labelled by $p$ (with $\Omega_1 = \Omega$ being largest). The upper limit for the sum over $p$ is 3 for a triple basis set.  The six nonlinear variational parameters $\alpha_p$, $\beta_p$ are determined by minimizing the energy on a six-dimensional energy surface.  This is accomplished by calculating analytically the derivatives $\partial E/\partial \alpha_p$, $\partial E/\partial \beta_p$ \cite{DrakeMakowski88} and finding their zeros by Newton's method.  The optimization produces a natural separation of the nonlinear parameters into short-range, intermediate, and asymptotic long-range sectors \cite{Sati2024}. The basis set also includes the screened hydrogenic term
$\psi_{1s}({\bf r}_1;Z)\psi_{np}({\bf r}_2;Z-1) \pm$ exchange for effective nuclear charges $Z$ and $Z-1$ respectively.

The rationale for improved accuracy is that a point of diminishing returns (and numerical instability) is reached with increasing $\Omega$, and so doubling or tripling the basis set allows more terms to be added without $\Omega$ becoming excessively large.

 The principal computational step  to obtain nonrelativistic wave functions and energies is to solve the generalized eigenvalue problem
\begin{equation}
{\bf H}\Psi = \lambda{\bf O}\Psi
\end{equation}
where {\bf H} is the Hamiltonian matrix with matrix elements $\langle\phi_{i^\prime,j^\prime,k^\prime}|H|\phi_{i,j,k}\rangle$,  {\bf O} is the overlap matrix in the nonorthogonal basis set, and $\Psi$ is the column vector of basis function coefficients $c_{i,j,k}^{(p)}$.  The Hamiltonian in center-of-mass (CM) coordinates is given by (in atomic units with $4\pi \epsilon_0 = 1$)
\begin{equation}
\label{eq:H}
H = \frac{p_1^2 + p_2^2}{2\mu} +\frac{{\bf p}_1\bdot {\bf p}_2}{M}
-\frac{Ze^2}{r_1} - \frac{Ze^2}{r_2} + \frac{e^2}{r_{12}}
\end{equation}
where ${\bf p}_1\bdot{\bf p}_2/M$ term is the mass polarization operator resulting from the motion of the nucleus in the
CM frame. The correction due to the finite nuclear mass is obtained by comparison with the infinite nuclear mass case $M = \infty$.  The power method is used to find the eigenvalue closest to an initial guess.  The strategy is to perform a sequence of calculations with increasing values of $\Omega$ and assess the rate of convergence to determine the accuracy.

As discussed previously \cite{Bondy2025}, the actual basis sets used in the calculations have two important modifications.  First, for the special case of $P$-states, the rate of convergence can be improved for the finite nuclear mass case by including a set of terms with the role of the $s$- and $p$-electrons interchanged. This is because the mass polarization operator has nonvanishing matrix elements of the form $\langle s(1)p(2)|{\bf p}_1\bdot{\bf p}_2/M|p(1)s(2)\rangle$ in the single-particle limit.  Second, various truncations can be judiciously introduced to reduce the overall redundancy in the triple basis set \cite{Bondy2025}.   

Table I shows a typical convergence pattern for the $27\,^3P$ state for $\Omega$ in the range 26 to 34.  $D(\Omega)$ denotes the difference between successive calculations, and $R(\Omega)$ is the ratio of successive differences.  There is evidently and even-odd alternation in the ratios, with an additional significant figure of accuracy with every increase of $\Omega$ by 2.  The extrapolated value, denoted by $\Omega=\infty$, is obtained by extending the ratios indefinitely.  The convergence pattern is similar for the case of finite nuclear mass. This degree of accuracy for the energy is much more than what is needed in comparison with experiment, but other quantities typically converge much more slowly with to no more than half as many significant figures.

\begin{table}[tb]
\caption{Convergence table for the nonrelativistic energy of
the $27\,^3P_1$ state assuming infinite nuclear mass.  $D(\Omega) = [E(\Omega) - E(\Omega-1)]\times10^{23}$ is the 
difference of successive entries, and 
$R(\Omega) = D(\Omega-1)/D(\Omega)$ is their ratio.  $\Omega = (i+j+k)_{\rm max}$ for each Pekeris shell, and $N$ is
the number of terms in the triple basis set.}
\begin{ruledtabular}
\begin{tabular}{lllrc}
\multicolumn{1}{c}{$\Omega$}   &\multicolumn{1}{c}{$N$} & \multicolumn{1}{c}{$E(\Omega)$ (E$_{\rm h}$)} &$D(\Omega)$& $R(\Omega)$ \\
\hline
26  & 3168  &$  -2.00068\,93526\,23570\,97339\,495  $&        &      \\                     
27  & 3508  &$  -2.00068\,93526\,23570\,97591\,082  $&251587  &      \\  
28  & 3892  &$  -2.00068\,93526\,23570\,97620\,998  $& 29916  &  8.41\\  
29  & 4284  &$  -2.00068\,93526\,23570\,97647\,677  $& 26679  &  1.12\\  
30  & 4724  &$  -2.00068\,93526\,23570\,97652\,802  $&  5125  &  5.21\\  
31  & 5172  &$  -2.00068\,93526\,23570\,97655\,186  $&  2384  &  2.15\\  
32  & 5672  &$  -2.00068\,93526\,23570\,97655\,585  $&   399  &  5.98\\  
33  & 6180  &$  -2.00068\,93526\,23570\,97655\,816  $&   231  &  1.72\\  
34  & 6744  &$  -2.00068\,93526\,23570\,97655\,851  $&    34  &  6.77\\  
$\infty$&   &$  -2.00068\,93526\,23570\,97655\,876(3)$&               \\
\end{tabular}
\end{ruledtabular}
\end{table}

 Table II presents a representative sample of comparisons between the nonrelativistic energies from the present work and previous calculations.  The most accurate for states up to $n=4$ are the all-exponential basis sets of Aznabaev et al.\cite{Aznabaev2018} with 22000 terms.  Next in 
accuracy are the present calculations up to $n=35$ with much smaller basis sets. A uniform standard of accuracy of at least 21 significant figures is maintained by progressively increasing the size of the basis set (as controlled by $\Omega$) from less than 3000 terms below $n=10$ to 7948 at
$n = 35$.  In contrast, the correlated B-spline calculations \cite{Chi2025} require up to 64000 terms with declining accuracy beyond $n = 20$. The advantage of the B-spline method is that it has better numerical stability, and a lengthy optimization of nonlinear parameters is avoided.  In most cases, 
the agreement with the present work is within the estimated uncertainty.  Exceptions are the $2\,^1P$ state where there is an 8$\sigma$
discrepancy, and a smaller discrepancy for the $3\,^1P$ state. A comprehensive tabulation of variational upper bounds and extrapolations for all $P$-states up to $n=35$ is given in the Appendix. 

\begin{table*}[tb]
\caption{Comparison of the nonrelativistic energies for the $n\,^1P$ and $^3P$ states of helium for the case of infinite nuclear mass.}
\begin{ruledtabular}
\begin{tabular}{rrrll}
\multicolumn{1}{c}{$n$}   &\multicolumn{1}{c}{$N$}&Reference & \multicolumn{1}{c}{$E(1snp\;^1P)$ } &
\multicolumn{1}{c}{$E(1snp\;^3P)$} \\
\hline
2 &22000&Aznabaev \cite{Aznabaev2018}   &$ -2.12384\,30864\,98101\,35924\,73331\,42374  $&$    -2.13316\,41907\,79283\,20514\,69927\, 63806 $\\          
  &2997 &Present work  &$ -2.12384\,30864\,98101\,35924\,1(22)\,                        $&$    -2.13316\,41907\,79283\,20514\,0(8) \,       $\\
  &64000&Chi \cite{Chi2025}  &$ -2.12384\,30864\,98101\,343(2)     \,                   $&$    -2.13316\,41907\,79283\,204(2)     \,        $\\
3 &22000&Aznabaev \cite{Aznabaev2018}  &$ -2.05514\,63620\,91943\,53692\,83410\,921     $&$    -2.05808\,10842\,74275\,33134\,26965\, 47203 $\\       
  &2617 &Present work  &$ -2.05514\,63620\,91943\,53692\,7(34)\,                        $&$    -2.05808\,10842\,74275\,33133\,9(33)\,       $\\
  &64000&Chi \cite{Chi2025} &$ -2.05514\,63620\,91943\,533(2)     \,                                     $&$    -2.05808\,10842\,74275\,3312(2)    \,        $\\
4 &22000&Aznabaev \cite{Aznabaev2018}  &$ -2.03106\,96504\,50240\,71475\,89314\,360941  $&$    -2.03232\,43542\,96630\,33195\,38824\, 67103 $\\         
  & 2775&Present work  &$ -2.03106\,96504\,50240\,71474\,2(12)\,                        $&$    -2.03232\,43542\,96630\,33195\,6(9) \,       $\\
  &64000&Chi \cite{Chi2025}   &$ -2.03106\,96504\,50240\,713(2)     \,                  $&$    -2.03232\,43542\,96630\,3319(2)    \,        $\\
6 & 2622&Present work  &$ -2.01383\,39796\,71740\,10238\,6(9) \,                        $&$    -2.01420\,95877\,37505\,91219(6)   \,        $\\
  &64000&Chi \cite{Chi2025}   &$ -2.01383\,39796\,71740\,1020(2)    \,                  $&$    -2.01420\,95877\,37505\,9133(4)    \,        $\\
9 & 2864&Present work  &$ -2.00615\,63846\,52853\,81834\,6(5) \,                        $&$    -2.00626\,72673\,66409\,03258\,0(14)\,       $\\
  &64000&Chi \cite{Chi2025}   &$ -2.00615\,63846\,52853\,8182(2)    \,                  $&$    -2.00626\,72673\,66409\,03258(2)   \,        $\\
12& 4868&Present work  &$ -2.00346\,52527\,04885\,79826\,035(5\,)                       $&$    -2.00351\,20065\,35142\,74556\,021(1\,8)     $\\
  &64000&Chi \cite{Chi2025}   &$ -2.00346\,52527\,04885\,7982(2)    \,                  $&$    -2.00351\,20065\,35142\,7456(2)    \,        $\\
15& 6160&Present work  &$ -2.00221\,86471\,04088\,30158\,826(6\,)                       $&$    -2.00224\,25712\,22150\,32643\,884(5\,)      $\\
  &64000&Chi \cite{Chi2025}   &$ -2.00221\,86471\,04088\,3016(4)    \,                  $&$    -2.00224\,25712\,22150\,3264(2)    \,        $\\
18& 5518&Present work  &$ -2.00154\,11387\,60913\,62470\,8318(\,8)                      $&$    -2.00155\,49769\,40232\,72665\,182(4\,)      $\\
  &64000&Chi \cite{Chi2025}   &$ -2.00154\,11387\,60913\,625(4)     \,                  $&$    -2.00155\,49769\,40232\,727(4)    \,         $\\
21& 6744&Present work  &$ -2.00113\,24817\,33017\,52951\,738(2\,)                       $&$    -2.00114\,11926\,56477\,67869\,1390(\,23)    $\\ 
  &64000&Chi \cite{Chi2025}   &$ -2.00113\,24817\,33017\,53(2)\,      \,                $&$    -2.00114\,11926\,56477\,68(2)\,      \,      $\\
24& 6744&Present work  &$ -2.00086\,71808\,46170\,11128\,223(6\,)                       $&$    -2.00087\,30145\,66616\,65939\,240(9\,)      $\\    
  &64000&Chi \cite{Chi2025}   &$ -2.00086\,71808\,46170\,1(2) \,      \,                $&$    -2.00087\,30145\,66616\,7(2) \,      \,      $\\
27& 6744&Present work  &$ -2.00068\,52565\,28882\,40212\, 1453(\,18)                    $&$    -2.00068\,93526\,23570\,97655\,876(3\,)      $\\
  &64000&Chi \cite{Chi2025}   &$ -2.00068\,52565\,2888(2)      \,      \,               $&$    -2.00068\,93526\,2357(2)  		\,    \,    $\\
30& 6744&Present work  &$ -2.00055\,51074\,62372\,97425\,91(24)                         $&$    -2.00055\,80928\,35757\,97528\,32(5)         $\\
33& 7948&Present work  &$  -2.00045\,88001\,04867\,46785\,90(2)                         $&$    -2.00046\,10426\,27369\,81858\,54(4)         $\\            
35& 7948&Present work  &$  -2.00040\,78810\,08092\,82096\,8(2)                           $&$    -2.00040\,97604\,35014\,70404\,5(2)          $\\                                                    \end{tabular}      
\end{ruledtabular}  
\end{table*}        

\section{Relativistic and QED Corrections}
\label{relativistic}
With the nonrelativistic wave functions in hand, it is a straightforward matter to calculate the relativistic and QED corrections as expectation values using the standard methods of nrQED.  Since all these terms decrease approximately in proportion to $1/n^3$, they are strongly suppressed for $n=24$ and higher, relative to the low-lying states. To order $\alpha^2$ Ry ($\alpha^4mc^2$), the leading terms are the well-known Breit interaction terms (superscripts denote the power of $\alpha$)
(in atomic units) \cite{DrakeYan92,Bethe}
\begin{equation}
E_{\rm rel}^{(2)} = \sum_{i=1}^5 \langle B_i\rangle + \frac{\mu}{M}(\tilde\Delta_2 + \tilde\Delta_{3Z})
\end{equation}
where $B_1 = -(p_1^4+p_2^4)/(8m^3c^2)$ arises from the relativistic variation of mass with velocity, $B_2$ is the orbit-orbit interaction defined by
\begin{equation}
B_2 = -\frac{e^2}{2(mc)^2}\left[\frac{{\bf p}_1\bdot {\bf p}_2}{r_{12}} + \frac{{\bf r}_{12}\bdot({\bf r}_{12}\bdot{\bf p}_1){\bf p_2}}{r_{12}^3}\right]
\end{equation}
$B_3 = B_{\rm so} + B_{\rm soo}$ contains the spin-orbit and spin-other-orbit interactions defined by
\begin{eqnarray}
B_{\rm so} &=& \frac{Z\mu_{\rm B}e(1+2a_{\rm e})}{mc}\left[\frac{{\bf r}_1\times{\bf p}_1\bdot{\bf s}_1}{r_1^3} + \frac{{\bf r}_2\times{\bf p}_2\bdot{\bf s}_2}{r_2^3}\right]\\
B_{\rm soo} &=& -\frac{\mu_{\rm B}e}{2mcr_{12}^3}{\bf r}_{12}\times\left[(3+4a_{\rm e}){\bf p}_-\bdot{\bf s}_+ -
{\bf p}_+\bdot{\bf s}_-\right]
\end{eqnarray}
where ${\bf p}_\pm = {\bf p}_1\pm{\bf p_2}$, ${\bf s}_\pm = {\bf s}_1\pm{\bf s_2}$, $\mu_{\rm B} = e\hbar/(2mc)$ is the Bohr magneton, and $a_{\rm e} \simeq\alpha/(2\pi) -0.328\,429\alpha^2$ is the electron anomalous magnetic moment.
$B_5$ is the spin-spin term $B_{\rm ss}$ defined by
\begin{equation}
B_{\rm ss}5 = 4\mu_e^2\left[\frac{8\pi}{3}\delta(r_{12}){\bf s}_1\bdot{\bf s}_2+
\frac{{\bf s}_1\bdot{\bf s}_2}{r_{12}^3} - \frac{3({\bf s}_1\bdot{\bf r}_{12}){\bf s}_2\bdot{\bf r}_{12}}{r_{12}^5}\right],
\end{equation}
where $\mu_e = \mu_{\rm B}(1 + a_e)$, and the relativistic recoil terms are
\begin{eqnarray}
\tilde\Delta_2&=&-\frac{Z e^2}{2(mc)^2} \sum_{j=1} ^2 \left[\frac{1}{r_j}{
\bf{p_+}} \bdot {\bf p}_j +\frac{1}{r_j^3}{\bf{r}}_j \bdot ({\bf{r}}
_j \bdot {\bf{p_+}}){\bf p}_j \right],\\
\label{eq:009}
\tilde\Delta_{3Z}&=&\frac{2Z\mu_{\rm B}e}{mc}\sum_{i=1} ^2 \frac{1}{r_i^3}{\bf{r}}
_i \times {\bf{p_+}} \bdot {\bf{s}}_i.
\label{eq:010}
\end{eqnarray}
  Finally
 $B_4$ contains the $\delta$-function terms $B_4 = \alpha^2\pi[\frac12\delta(r_1)
+ \frac12\delta(r_2) - \delta(r_{12})]$.  To each of these, there are finite nuclear mass corrections of order $\alpha^2\mu/M$ arising from (i) the mass scaling of each term, and (ii) cross terms between the $B_i$ terms in $E_{\rm rel}^{(2)}$ and the mass polarization operator in Eq.\ (\ref{eq:H}).
The relativistic recoil terms $\tilde\Delta_2$ and $\tilde\Delta_{3Z}$ arise from the transformation of the Breit interaction to CM plus relative coordinates \cite{DrakeYan92,Stone1963}.
In the material to follow, the mass scaling and recoil terms taken together are denoted by $E_{\rm RR,M}^{(2)}$, and the cross terms (ii) by $E_{\rm RR,X}^{(2)}$.

The leading QED corrections of order $\alpha^3$ Ry ($\alpha^5mc^2)$ due to electron self-energy and vacuum polarization can be divided into an electron-nucleus part ($E_{\rm L,1}^{(3)}$) \cite{Kabir} and an electron-electron part ($E_{\rm L,2}^{(3)}$) \cite{Araki,Sucher} defined by
\begin{eqnarray}
 E_{\rm L,1}^{(3)} =&&\!\!\!\!\! \textstyle\case43Z\alpha^3[\case{19}{30} -\ln(Z\alpha)^2 - \ln k_0]\langle\delta(r_1) + \delta(r_2)\rangle\;\;\\
 E_{\rm L,2}^{(3)} =&&\!\!\!\!\! \textstyle\alpha^3[(\case{89}{15} + \case{14}{3}\ln\alpha - \case{20}{3}{\bf s}_1\bdot{\bf s}_2)\langle\delta(r_{12})\rangle-\case{14}{3}Q]
\end{eqnarray}
where $k_0$ is Bethe's mean excitation energy \cite{Bethe} in units of $Z^2$ Ry.
The advantage of subtracting out the $\ln Z^2$ scaling of the Bethe logarithm and including it instead in the $\ln(Z\alpha)^2$ term is that the value is then close to  the hydrogenic value 2.984\,128\,556 ($\pm$0.5\%) for all states of all light atoms and ions studied \cite{YanDrake2003,PachuckiKomasa2004,Korobov2019,Lesiuk2024}.  Its value for the $27\,^1P$ and $^3P$ states of helium can be accurately estimated from the $1/n$ expansions of Drake \cite{Drake2001} or Korobov \cite{Korobov2019} with the result
\begin{eqnarray*}
\ln k_0(27\,^1P) &=& 2.984\,128\,3136(3)  -0.000\,001\,25(7)\mu/M\\
\ln k_0(27\,^3P) &=& 2.984\,128\,2191(10)  +  0.000\,003\,49(9)\mu/M
\end{eqnarray*}
which is very close to the hydrogenic value. 
 For $\Delta E_{\rm L,2}^{(3)}$, the $Q$ term
is defined by the improper integral
\begin{equation}
\label{Q0}
Q = \frac{1}{4\pi}\lim_{\epsilon\rightarrow 0}\langle
r_{12}^{-3}(\epsilon)+ 4 \pi (\gamma +\ln \epsilon) \delta({\bf r}_{12})\rangle\,
\end{equation}
where $\epsilon$ is the radius of an infinitesimal sphere that is excluded from the range of integration, and $\gamma$ is Euler's constant.
\begin{table*}[tb]
\caption{Evaluation of dominant and remainder terms of $O(\alpha^4)$ and $O(\alpha^5)$ Ry for the $2\,P$ states of helium for
$1/n^3$ scaling of the remainder.  Units are MHz. }
\label{scaling}
\begin{ruledtabular}
\begin{tabular}{lddddd}
Term   &\multicolumn{1}{r}{$2\;^1P_1$}&\multicolumn{1}{r}{$2\;^3P_0$} &\multicolumn{1}{r}{$2\;^3P_1$} &
\multicolumn{1}{r}{$2\;^3P_2$} &\multicolumn{1}{r}{$2\;^3P_{\rm c}$}\\
\hline
                     & \multicolumn{5}{c}{Partial sum of dominant terms}\\
$E_{\rm anom}^{(4)}$           &0.0000  & -0.0339&   0.0178&  -0.0004&    0.0000\\
$E_{\rm R1}$                   &1.6634  &-20.6548& -20.6548& -20.6548&  -20.6548\\
$E_{\rm R2}$                   &0.0148  & -0.1842&  -0.1842&  -0.1842&   -0.1842\\
$E_{\rm e1} + E_{\rm e2}$      &0.3246  &  0.0000&   0.0000&   0.0000&    0.0000\\
$E_{{\rm st},2}^{(4)}$  &4.7550  &  0.0000&  -4.7550&   0.0000&   -1.5850\\
$\alpha^4\ln(\alpha)$          &0.2119  &  0.0000&   0.0000&   0.0000&    0.0000\\
Subtotal                       &6.9697  &-20.8729& -25.5762& -20.8430&  -22.4240\\
                     & \multicolumn{5}{c}{Complete $O(\alpha^4)$ totals for comparison} \\
$E_{\rm c}^{(4)}$ \cite{Pachucki2017} &8.8180  &-21.8320& -21.8320& -21.8320&  -21.8320\\
$\Delta E_J^{(4)}$ \cite{Patkos2021}  &0.0000  & -4.9695&  -3.4523&   3.0653&    0.0000\\
Total                          &8.8180  &-26.8015& -25.2843& -18.7667&  -21.8320\\
                               &        &        &         &         &          \\
Difference                     &1.8483  & -5.9286&  0.2919 &   2.0763&    0.5921\\
$\alpha^5$                     &0.810   &  0.223 &  0.223  &   0.223 &    0.223 \\
Total = $E_{\rm rmdr}$  &2.658   & -5.705 &  0.068  &   1.853 &    0.369 \\
\end{tabular}      
\end{ruledtabular}  
\end{table*}

Contributions of order $\alpha^4$ Ry are more difficult to evaluate.  They have been
calculated in their entirety for the $2\,^1P_1$ state by Pachucki et al.\
\cite{Pachucki2017}.  Our strategy is to calculate the dominant parts that can be easily evaluated, subtract these from the results in Ref.\ \cite{Pachucki2017}, and estimate the remainder from its approximately $1/n^3$ scaling with $n$. For example, for the $2\,^1P_1$ state, the total contribution of order $\alpha^4$ is 8.818 MHz \cite{Pachucki2017}.  The largest contribution comes from the
$2\;^1P_1 - 2\;^3P_1$ second-order singlet-triplet mixing term induced by the spin-dependent terms in the Breit interaction acting twice of the form
\begin{equation}
E_{\rm st}^{(4)} = \sum_{n'=2}^{\infty}\frac{\langle 2\,^1P_1 | B |n'\,^3P_1\rangle\langle n'\,^3P_1| B |2\,^1P_1\rangle}{E(2\,^1P_1) - E(n'\,^3P_1)}\,,
\end{equation} 
including an integration over the continuum.  The $n'=2$ term $E_{\rm st,2}^{(4)}$ is largest (4.7549 MHz) \cite{Drake-Long}, since the energy denominator is the smallest.  In general, it is the $n'=n$ term that is largest for the $n\,^1P_1$ state, denoted by $E_{{\rm st},n}^{(4)}$, and equal but opposite sign for the $n\,^3P_1$ state.  

There are also large radiative terms that come from a sum of one- and two-loop contributions denoted by $E_{\rm R1}$ and $E_{\rm R2}$ in Ref.\ \cite{Pachucki2006-P} (1.995 MHz), electron-electron radiative terms $E_{\rm e1}$ and $E_{\rm e2}$ (0.3246 MHz), and an $\alpha^4\ln\alpha$ term (0.2120 MHz) \cite{Drake-Khrip}.  The sum of these terms is 6.9697 MHz, leaving a remainder $E^{(4)}_{\rm rmdr}$ of $8.818 - 6.9697 = 1.848$ MHz that comes from a
mixture of first- and second-order terms that are much more difficult to calculate, as
discussed in detail in Refs.\ \cite{Pachucki2006-P,Pachucki2006-S}.  However, all these terms scale roughly in proportion to $1/n^3$, so one can expect a corresponding contribution of approximately $1.848(2/27)^3 = 0.75$ kHz for the $27\,^1P_1$ state. These and the corresponding terms for the triplet states are summarized in Table \ref{scaling}, along with a similar term of order $\alpha^5$ \cite{Pachucki2017}.  The estimated remainders $E^{(4)}_{\rm rmdr}$ are included in the final results with the entire amount taken as the uncertainty. As a check, the scaling argument applied to the singlet-triplet mixing term $E^{(4)}_{{\rm st},n}$ reduces its value by a factor of 2460 for $n=27$ to 1.93 kHz, which is close to the correct value 1.597 kHz (see Table \ref{n=27}).  

\begin{table*}[!tb]
\caption{ Contributions to the $27\,^1P_1$ and $27\,^3P_J$ negative ionization frequencies of $^4$He. $E_{\rm radiative}^{(4)}$ denotes
the sum of radiative terms $E_{\rm R1} + ER_{\rm R2} + E_{\rm e1} + E_{\rm e2}$, and $E_{\rm rmdr}^{(4)}$ denotes
the remainders from Table \ref{scaling} scaled by a factor of $(2/n)^3$. Units are MHz.}
\label{n=27}
\begin{ruledtabular}
\begin{tabular}{l c dddd}
\multicolumn{1}{c}{Contribution}& Order          &\multicolumn{1}{c}{$27\,^1P_1$}&\multicolumn{1}{c}{$27\,^3P_0$}&
\multicolumn{1}{c}{$27\,^3P_1$}&\multicolumn{1}{c}{$27\,^3P_2$}\\
\hline
Nonrelativistic $E_{\rm nr}$                &                &-4508153.32664   & -4535100.64070 &-4535100.64070  &-4535100.64070\\
1st.\ order mass pol.\ $E_{\rm M,1}$        & $\mu/M$        &      18.80897   &      -20.91429 &     -20.91429  &     -20.91429\\
2nd.\ order mass pol.\ $E_{\rm M,2}$        & $(\mu/M)^2$    &      -0.08590   &       -0.08591 &      -0.08591  &      -0.08591\\
Relativistic $E_{\rm rel}^{(2)}$            & $\alpha^2$     &      -9.36849   &       10.02726 &      -0.01344  &      -0.83985\\
Relativ.\ finite mass $E_{\rm RR,M}^{(2)}$  &$\alpha^2\mu/M$ &      -0.00478   &       -0.00076 &       0.00142  &       0.00049\\
Relativistic recoil $E_{\rm RR,X}^{(2)}$    &$\alpha^2\mu/M$ &       0.00262   &        0.00380 &       0.00200  &       0.00185\\
Anomalous mag.\ mom.                   &$\alpha^3 + \alpha^4$&       0.0       &        0.01160 &      -0.00661  &       0.00165\\
$E_{\rm QED}^{(3)}$                         & $\alpha^3$     &       0.02143   &       -0.44721 &      -0.44721  &      -0.44721\\
Singlet-triplet mixing $E_{\rm st,27}^{(4)}$& $\alpha^4$     &       0.00160   &        0.00000 &      -0.00160  &       0.00000\\
$E_{\rm radiative}^{(4)}$                   & $\alpha^4$     &       0.00107   &       -0.00762 &      -0.00762  &      -0.00762\\
$E_{\rm rmdr}^{(4)}   $     & $\alpha^4$    &       0.0011(11)&       -0.0023(23)&     0.0010(10)&       0.00013(13)\\ 
$E_{\rm QED}^{(5)}$                         & $\alpha^5$     &      -0.0000    &        0.00075 &       0.00075  &       0.00074\\
Finite nuclear size $E_{\rm nuc}$           & $(R/a_0)^2$    &       0.00003   &       -0.00029 &      -0.00029  &      -0.00029\\
Total                                       &                &-4508143.9491(11)& -4535112.0557(23)&-4535122.1125(10)&-4535122.9310(7)\\
Center of gravity                           &                &-4508143.9491(11)&                &-4535121.4498(11)&               \\
Experiment$^{\,a}$                                  &                &-4508143.903(78) &                &-4535121.435(20)  &              \\
Difference                                  &                &       0.046(78) &                &       0.014(20)  &              \\
%
\end{tabular}
\end{ruledtabular}\\ 
\hbox to \textwidth{$^{a\,}$From quantum defect fit \cite{Merkt2025}.\hfill}
\end{table*}

The last correction to be included is that due to finite nuclear size.  To lowest order, the energy shift to sufficient accuracy is $E_{\rm nuc} = \frac23\pi Z(R/a_0)^2\langle\delta({\bf r}_1) + \delta({\bf r}_2)\rangle$ where $a_0$ is the Bohr radius and $R=1.6786(12)$ fm \cite{Pachucki2024} is the radius of the nuclear charge distribution for $^4$He.

As an example, the various calculated contributions to the ionization energy of the $27\,^1P_1$ and $27\,^3P_J$ states of $^4$He relative to $^4$He$^+(1s)$ are summarized in Table \ref{n=27}. For every entry, the corresponding hydrogenic value for the $^4$He$^+(1s)$ ion is subtracted before converting from a.u.\ to MHz. The dominant source of uncertainty ($\pm$1 kHz) is the residual QED contribution of order $\alpha^4$ Ry.  However, this level of theoretical uncertainty is already much better than for the low-lying states of helium, and it establishes an absolute point of reference for transitions to low-lying states. Table \ref{alln} summarizes the calculated ionization energies for all the $n\,^1P_1$ and $n\,^3P_c$ (center of gravity) states of helium for 
$2\le n \le 35$, and Table \ref{fine} lists separately the calculated fine-structure energy shifts relative to the center of gravity.  As in Table \ref{n=27}, all the
uncertainties are the entire amount of the residual QED contribution $E^{(4)}_{\rm rmdr}$ of order $\alpha^4$ Ry.  The results for $n = 2$ 
are known much more accurately from the calculations of Patk\'o\v{s} et al.\ \cite{Pachucki2017,Patkos2021}.  They are included here to provide a 
consistent set of calculations over the entire range of $n$.  The results for $n = 3$ and higher are the most accurate in the literature.  Tables of matrix elements for other $n\,P$ states are available at \cite{data}.

\begin{table}[!tb]
\caption{Calculated negative ionization frequencies of the $n\;^1P_1$ and $n\;^3P_{\rm c}$ states of $^4$He.  Units are MHz.}
\label{alln}
\begin{ruledtabular}
\begin{tabular}{rdd}
$n$ & I(n\;^1P_1)/h &I(n\;^3P_{\rm c})/h\\
\hline   
2  &  -814\,709\,146.5(2.7)  &  -876\,106\,246.6(2.3)    \\
3  &  -362\,787\,967.3(8)    &  -382\,117\,481.1(7)      \\
4  &  -204\,397\,210.37(33)  &  -212\,661\,130.22(29)    \\
5  &  -130\,955\,541.64(17)  &  -135\,204\,995.78(15)    \\
6  &   -91\,009\,810.41(10)   &  -93\,472\,929.86(8)     \\
7  &   -66\,901\,127.43(6)    &  -68\,453\,141.74(5)     \\
8  &   -51\,242\,587.32(4)    &  -52\,282\,461.860(35)   \\
9  &   -40\,501\,246.341(29)  &  -41\,231\,541.911(24)   \\  
10 &   -32\,814\,665.275(21)  &  -33\,346\,972.326(18)   \\
11 &   -27\,125\,436.770(16)  &  -27\,525\,292.385(13)   \\
12 &   -22\,797\,031.709(12) &   -23\,104\,960.774(10)   \\
13 &   -19\,427\,674.895(10) &   -19\,669\,820.901(8)    \\
14 &   -16\,753\,624.158(8)  &   -16\,947\,462.118(6)    \\
15 &   -14\,595\,939.724(6)  &   -14\,753\,507.979(5)    \\
16 &   -12\,829\,749.819(5)  &   -12\,959\,559.3773(37)  \\
17 &   -11\,365\,768.212(4)  &   -11\,473\,973.5454(29)  \\
18 &   -10\,138\,783.8411(36)&   -10\,229\,924.3219(23)  \\
19 &    -9\,100\,271.0054(31)&    -9\,177\,753.8936(19)  \\ 
20 &    -8\,213\,516.6237(27)&    -8\,279\,939.6243(15)  \\ 
21 &    -7\,450\,330.3493(23)&    -7\,507\,701.8588(12)  \\ 
22 &    -6\,788\,776.0678(20)&    -6\,838\,668.5736(12)  \\ 
23 &    -6\,211\,577.8174(17)&    -6\,255\,236.6672(11)  \\ 
24 &    -5\,704\,980.3477(15)&    -5\,743\,402.1284(11)  \\ 
25 &    -5\,257\,921.9483(14)&    -5\,291\,911.8055(11)  \\ 
26 &    -4\,861\,425.4377(12)&    -4\,891\,639.5603(10)  \\ 
27 &    -4\,508\,143.9491(11)&    -4\,535\,121.4498(10)  \\ 
28 &    -4\,192\,018.1260(10)&    -4\,216\,205.2379(10)  \\
29 &    -3\,908\,014.5582(9) &    -3\,929\,783.1919(10)  \\ 
30 &    -3\,651\,924.1759(8) &    -3\,671\,586.2926(10)  \\ 
31 &    -3\,420\,205.3943(7) &    -3\,438\,024.2366(9)   \\ 
32 &    -3\,209\,861.0067(6) &    -3\,226\,059.9508(9)   \\ 
33 &    -3\,018\,340.7741(6) &    -3\,033\,110.3603(9)   \\ 
34 &    -2\,843\,463.7608(5) &    -2\,856\,967.3194(9)   \\ 
35 &    -2\,683\,355.9699(5) &    -2\,695\,734.1549(9)   \\
\end{tabular}
\end{ruledtabular}\\
\end{table}

\begin{table}[!tb]
\caption{Calculated fine structure shifts of the  $n\;^3P_J$ states of $^4$He, relative to the center of gravity.  Units are MHz.}
\label{fine}
\begin{ruledtabular}
\begin{tabular}{rddd}
$n$ &\multicolumn{1}{c}{$\Delta I(n\;^3P_0)/h$} &\multicolumn{1}{c}{$\Delta I(n\;^3P_{\rm 1})/h$}&\multicolumn{1}{c}{$\Delta I(n\;^3P_{\rm 1})/h$}\\
\hline   
 2  &  27599.1(2.3)    &     -2017.9(2.3)    &     -4309.1(2.3)   \\
 3  &   7577.8(7)      &      -535.5(7)      &     -1194.3(7)     \\
 4  &   3089.02(29)    &      -217.69(29)    &      -487.19(29)   \\
 5  &   1552.43(15)    &      -109.42(15)    &      -244.83(15)   \\
 6  &    888.15(8)     &       -62.63(8)     &      -140.05(8)    \\
 7  &    555.01(5)     &       -39.15(5)     &       -87.51(5)    \\
 8  &    369.767(35)   &       -26.093(35)   &       -58.298(35)  \\
 9  &    258.631(24)   &       -18.254(24)   &       -40.774(24)  \\
 10 &    187.941(18)   &       -13.267(18)   &       -29.628(18)  \\
 11 &    140.844(13)   &        -9.943(13)   &       -22.203(13)  \\
 12 &    108.262(10)   &        -7.644(10)   &       -17.066(10)  \\
 13 &     85.004(8)    &        -6.002(8)    &       -13.400(8)   \\
 14 &     67.961(6)    &        -4.799(6)    &       -10.713(6)   \\
 15 &     55.187(5)    &        -3.897(5)    &        -8.699(5)   \\
 16 &     45.4239(37)  &        -3.2074(37)  &        -7.1604(37) \\
 17 &     37.8350(29)  &        -2.6715(29)  &        -5.9641(29) \\
 18 &     31.8470(23)  &        -2.2486(23)  &        -5.0203(23) \\
 19 &     27.0590(19)  &        -1.9104(19)  &        -4.2655(19) \\ 
 20 &     23.1847(15)  &        -1.6368(15)  &        -3.6549(15) \\ 
 21 &     20.0164(12)  &        -1.4134(12)  &        -3.1553(12) \\ 
 22 &     17.3998(12)  &        -1.2282(12)  &        -2.7431(12) \\ 
 23 &     15.2203(11)  &        -1.0742(11)  &        -2.3995(11) \\ 
 24 &     13.3902(11)  &        -0.9450(11)  &        -2.1111(11) \\ 
 25 &     11.8421(11)  &        -0.8356(11)  &        -1.8671(11) \\ 
 26 &     10.5238(10)  &        -0.7425(10)  &        -1.6593(10) \\ 
 27 &      9.3941(10)  &        -0.6627(10)  &        -1.4812(10) \\ 
 28 &      8.4206(10)  &        -0.5939(10)  &        -1.3278(10) \\
 29 &      7.5770(10)  &        -0.5343(10)  &        -1.1948(10) \\ 
 30 &      6.8425(10)  &        -0.4824(10)  &        -1.0790(10) \\ 
 31 &      6.1999(9)   &        -0.4370(9)   &        -0.9778(9)  \\ 
 32 &      5.6353(9)   &        -0.3971(9)   &        -0.8888(9)  \\ 
 33 &      5.1373(9)   &        -0.3620(9)   &        -0.8103(9)  \\ 
 34 &      4.6963(9)   &        -0.3308(9)   &        -0.7408(9)  \\ 
 35 &      4.3043(9)   &        -0.3031(9)   &        -0.6790(9)  \\ 
\end{tabular}
\end{ruledtabular}\\
\end{table}

\section{Results}
\label{results}
A direct comparison with experiment for the $2\,^1S_0 - n\,^1P_1$ and $2\,^3S_1-n\,^3P_{\rm c}$ transition frequencies is complicated by the fact that the energies of the lower $S$-states is not known at this level of accuracy, and it is indeed the purpose of the experimental studies to determine it by use of quantum defect theory to extrapolate the series of measurements to the series limit.  We therefore present instead the derived ionization energies of the lower $2\,^1S_0$ and $2\,^3S_1$ states obtained by adding the negative theoretical ionization frequencies $-I(n\,^1P)$ from Table IV to the measured $2\,S - n\,P$ transition frequencies from Refs.\ \cite{Clausen2021} and \cite{Clausen2025}.  The results are shown in Table \ref{results} for the seven singlet measurements in the range $24\le n \le 35$, and the parallel four triplet measurements. In addition, the singlet measurements can be converted to triplet measurements of the $2\,^3S_1$ state by adding the highly accurate singlet-triplet transition frequency $\nu(2\,^1S_0 - 2\,^3S_1) =  192\,510\,702.14872$ MHz \cite{Rengelink2018}, resulting in 11 independent measurements of the $2\,^3S_1$ ionization energy.  The results are summarized in Fig.\ 1. First, the weighted average ionization energy of the $2\,^1S_0$ state (960\,332\,040.546(9) MHz) marginally agrees with the less accurately known theoretical value $960\,332 038.0 (1.9)$ MHz.  Second, the two sets of combined results for the $2\,^3S_1$ state (one from the singlets and one from the triplets) differ from each other by a similar amount (58 kHz from QDT exrapolation Ref.\ \cite{Clausen2025} and 30 kHz from the present work), so in this regard the results are consistent.  Although the systematic uncertainty on the singlet side is large ($\pm$20 kHz), it is clear from Fig.\ 1 that there is a statistically meaningful difference between the two sets of results. 

The sharpest comparison between the present theory and the QDT extrapolations is provided by the four triplet-$P$ measurements with $n = 27, 
29, 33,$ and 35.  The weighted average of the four measurements is 1152\,842\,742.7247(78) MHz, where the uncertainty of $\pm$0.0078 MHz is the rms  standard deviation of the four measurements divided by $\sqrt{3}$, and including the theoretical uncertainty added linearly.  The result is larger than the QDT value (1152\,842\,742.7082(55)$_{\rm stat}(25)_{\rm sys}$ MHz) by $16\pm10$ kHz.  The additional $\pm$ 20 kHz systematic uncertainty in the measurements is common to both sets of data and so is assumed to cancel.
Similarly, our grand average over the two sets of measurements, weighted in proportion to the inverse square of the total uncertainties for the singlet and triplet data, is $1152\,842\,742.722(11)$ MHz.  This agrees with, but lies $14\pm17$ kHz deeper than the similarly weighted Clausen QDT value 1152\,842\,742.707(13) kHz. Taken together, the overall good agreement confirms a 9$\sigma$ discrepancy of  $0.471\pm0.053$ MHz with the theoretical value $1152\,842\,742.231(52)$ MHz.  Further work on the theoretical side has not yielded any significant new results to resolve the discrepancy \cite{Yerokhin2022,Yerokhin2023}.  As discussed by Clausen et al.\ \cite{Clausen2025}, the discrepancy is even bigger (15$\sigma$) for the $2\,^3P$ state
as derived from the $2\,^3P_J - 3\,^3D_J$ transition frequency \cite{Luo2016}.  It is particularly significant that there is a nearly identical discrepancy of $0.482 \pm0.053$ kHz for the $^3$He isotope \cite{Clausen2025c}, despite the additional complications of hyperfine structure for this case.
\begin{figure}
\includegraphics[width=3.7in]{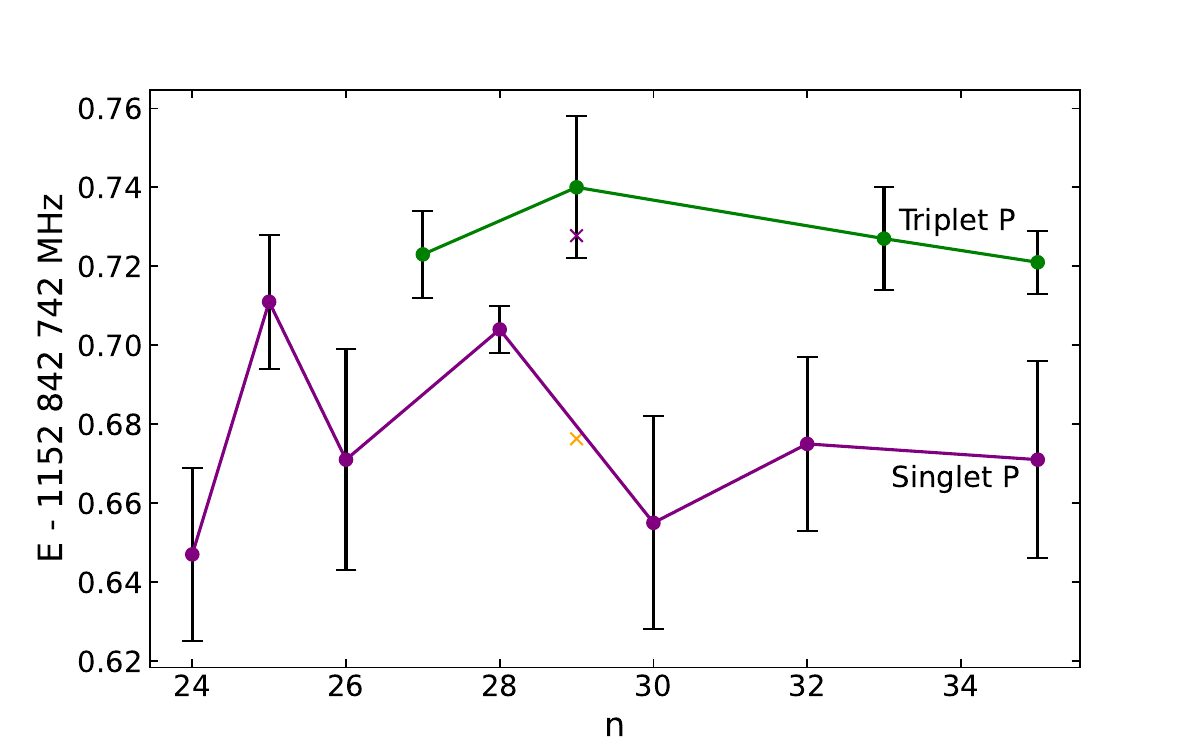}
\caption{Ionization energy of the $2\,^3S_1$ state of $^4$He as determined by adding the calculated ionization frequency of the $n\,^3P_{\rm c}$ state to the measured $2\,^3S_1-n\,^3P_{\rm c}$ transition frequency \cite{Clausen2025}, and similarly for the singlet case \cite{Clausen2021} together with the measured $2\,^1S_0-2\,^3S_1$ transition frequency \cite{Rengelink2018}.}
\label{ionization}
\end{figure} 

\section{Discussion}
In summary, we have completed an extensive set of high precision variational calculations for all singlet and triplet $P$-states of $^4$He up to $n=35$, while maintaining a uniform standard of accuracy of 1 part in $10^{22}$ for the nonrelativistic energy.  This is more than sufficient to determine the lowest-order $\alpha^2$ relativistic and $\alpha^3$ QED corrections at the 1 kHz level of accuracy. The dominant parts of the $\alpha^4$ Ry are calculated directly for all $n$.  The leading sources of uncertainty are the uncalculated terms of order $\alpha^4$ Ry. These are estimated by assuming that the accurately known values for $n=2$ \cite{Patkos2021} scale as $1/n^3$ with $n$, and taking the entire amount of the scaled value as the uncertainty.  This procedure reduces the uncertainty to about 1 kHz at $n = 24$. A complete calculation of these $\alpha^4$ terms for the Rydberg $P$-states is in progress and will be published separately.

The main significance of the results is that a direct comparison between theory and experiment for $n$ in the range $24 \le n \le 35$ provides 11 independent measurements of the ionization frequency of the controversial $2\,^3S_1$ state. The grand average over singlets and triplets yields a 9$\sigma$ disagreement between experiment and theory of $0.471\pm0.053$ MHz, in close agreement with the result obtained by quantum defect extrapolation to the series limit.  However, QDT theory itself is a semiempirical fitting procedure that is not derivable from fundamental atomic physics.  Although widely used and highly successful, it has never been tested at this high level of experimental accuracy.  Especially the Ritz procedure of including only even powers in the quantum defect expansion may not remain valid in the presence of relativistic corrections \cite{DrakeAdv}.  It is therefore significant that the two methods of determining the ionization energy of the $2\,^3S_1$ state agree at the level of $\pm$17 kHz from the Grand Average in Table \ref{results}.

Concerning theory for the $2\,^3S_1$ and $2\,^3P_1$ states, the calculation of Patk\'o\v{s} et al.\ \cite{Patkos2021} greatly reduced the uncertainty by a complete calculation of the QED terms of order $\alpha^5$ Ry.  They also estimated the contribution from the next higher-order $\alpha^6$ Ry term, based on a hydrogenic approximation. The result of --0.158(52) MHz for the (positive) ionization energy of the $2\;^3S_1$ state stands as the dominant source of theoretical uncertainty. In contrast, this uncertainty, and the entire $\alpha^4$ Ry term, are suppressed to the 1 kHz level or below by the $1/n^3$ scaling at $n=24$, thereby allowing a clear comparison with experiment for the singlet and triplet spectra ($\pm$95 kHz),  uncomplicated by higher-order QED uncertainties, or the use of quantum defect theory for extrapolations to the series limit. This leaves unexplained a 9$\sigma$ disagreement between theory and experiment for the ionization energy of the $2\,^3S_1$ state ($0.471\pm0.053$ MHz)

\begin{table*}[!tb]
\caption{Combined ionization energies $I^\prime$ of the $2\,^1S_0$ and $2\,^3S_1$ states of $^4$He obtained by adding the theoretical ionization energy $I_{\rm theo.}(n\,^1P_1)$ or $I_{\rm theo.}(n\,^3P_{\rm c})$ to the experimental $2\,^1S_0 - n\,^1P_1$ or $2\,^3S_1 - n\,^3P_{\rm c}$ transition frequencies respectively, and compared with experimental ionization energies obtained by quantum defect extrapolation. The statistical
and systematic uncertainties for the experimental data are shown in brackets. Units are MHz.}
\label{results}
\begin{ruledtabular}
\begin{tabular}{lllll}
\multicolumn{1}{l}{$n$}&$I_{\rm theo.}(n\;P)/h$ &$\nu_{\rm exp.}(2\;S-n\;P)$&\multicolumn{1}{c}{$I^\prime(2\;^1S_0)/h$}
&\multicolumn{1}{c}{$I^\prime(2\;^3S_1)/h$}\\
\hline   
\multicolumn{5}{c}{Singlets}                                                         \\
  24  &  5704\,980.3477(15)  & 954\,627\,060.151(22)  &    960\,332\,040.499(23)  & 1152\,842\,742.647(23)$^{\,\rm a}$ \\
  25  &  5257\,921.9483(14)  & 955\,074\,118.614(17)  &    960\,332\,040.562(18)  & 1152\,842\,742.711(18)$^{\,\rm a}$ \\
  26  &  4861\,425.4377(12)  & 955\,470\,615.084(28)  &    960\,332\,040.522(29)  & 1152\,842\,742.671(29)$^{\,\rm a}$ \\
  28  &  4192\,018.1260(10)  & 956\,140\,022.429(6)   &    960\,332\,040.555(7)   & 1152\,842\,742.704(7)$^{\,\rm a}$  \\
  30  &  3651\,924.1759(8)   & 956\,680\,116.330(27)  &    960\,332\,040.506(28)  & 1152\,842\,742.655(28)$^{\,\rm a}$ \\
  32  &  3209\,861.0067(6)   & 957\,122\,179.519(22)  &    960\,332\,040.526(23)  & 1152\,842\,742.675(23)$^{\,\rm a}$ \\
  35  &  2683\,355.9699(5)   & 957\,648\,684.552(25)  &    960\,332\,040.522(26)  & 1152\,842\,742.671(26)$^{\,\rm a}$ \\
Weighted average&       &                        &    960\,332\,040.547(9)   & 1152\,842\,742.694(10)$_{\rm stat}(20)_{\rm sys}$  \\
QDT Exp't.\ \cite{Clausen2021}&   &              &                           & 1152\,842\,742.650(32)$_{\rm stat}(20)_{\rm sys}$\\
Difference&             &                        &                  &\phantom{1152\,842\,74}0.044(34)$_{\rm stat}$\\
\multicolumn{5}{c}{Triplets}                                                        \\
  27  &  4535\,121.4498(10) & 1148\,307\,621.274(11)  &                       & 1152\,842\,742.723(12)                  \\
  29  &  3929\,783.1919(10) & 1148\,912\,959.549(17)  &                       & 1152\,842\,742.740(18)                  \\
  33  &  3033\,110.3603(9) & 1149\,809\,632.367(13)  &                       & 1152\,842\,742.727(14)                  \\
  35  &  2695\,734.1549(9) & 1150\,147\,008.567(8)   &                       & 1152\,842\,742.721(9)                   \\
Weighted average&     \,        &    \,    \,             &                       & 1152\,842\,742.7247(78)$_{\rm stat}(25)_{\rm sys}$\\
QDT Exp't.\ \cite{Clausen2025}&   &              &                       & 1152\,842\,742.7082(55)$_{\rm stat}(25)_{\rm sys}$\\
Difference&            &                         &              &\phantom{1152\,842\,74}0.016(10)$_{\rm stat}$     \\
\hbox to 0pt{Grand average}&       &    \,    \,             &           & 1152\,842\,742.721(11)                  \\
QDT Exp't.\ \cite{Clausen2021,Clausen2025}&   &  &                       & 1152\,842\,742.707(13)                  \\
Difference&            &                         &              &\phantom{1152\,842\,74}0.014(17)                  \\
\end{tabular}
\end{ruledtabular}\\
\hbox to \textwidth{$^{\rm a\,}$Obtained by adding $\nu(2\,^1S_0 - 2\,^3S_1) =  192\,510\,702.14872$ MHz \cite{Rengelink2018} to $I^\prime(2\;^1S_0)/h$.\hfill}
\end{table*}

\begin{acknowledgements}
This research was supported by the Natural Sciences and Engineering Research Council of Canada (NSERC) and by the Digital Research Alliance of Canada/Compute Ontario. The authors are grateful to Fr\'ed\'eric Merkt for helpful advice and correspondence concerning the treatment of uncertainties in the experimental data reported in Table \ref{results}. 

\end{acknowledgements}
\section*{Appendix}
This Appendix presents two supplementary tabulations.  First, Table \ref{bounds} presents a complete tabulation of the variational upper bounds and extrapolations for the nonrelativistic energies of the singlet 
and triplet $P$-states of helium with
prncipal quantum number $4\le n \le 35$. The number of terms $N$ in the triple basis set gradually increases in order to maintain a consistent level of accuracy of approximately 22 significant figures.  

Second, Table \ref{matrix} gives  matrix elements of the spin-dependent operators $B_{\rm so}^\prime$,
$B_{\rm soo}^\prime$, $B_{\rm ss}^\prime$, and $\Delta_{3Z}^\prime$, where, for simplicity of comparison,  the prime indicates that the anomalous magnetic moment correction is not included.  For each $n$, the first entry is the matrix element $\langle n\,^3P_1|B_{X,\infty}| n\,^3P_1\rangle$ for the case of infinite nuclear mass, and the second entry is the complete mass polarization correction $\langle n\,^3P_1|B_{X,\rm mp^\prime}| n\,^3P_1\rangle$ divided by $\mu/M$ for the case of $^4$He, as defined in Ref.\ \cite{Drake1992}.  It is therefore the first-order perturbation correction due to mass polarization, plus higher-order terms in $\mu/M$ summed to infinity. The complete matrix element for the case of finite nuclear mass then 
corresponds to the operator
\begin{equation}
  B_{X}^\prime(M) = (1-\mu/M)^3[B_{X,\infty}^\prime + (\mu/M) B_{X,\rm mp}^\prime]
 \end{equation}
 For example, with $M/m_{\rm e} = 294.299\,541\,71(17)$ \cite{Codata}, and $\mu/M = 1/(1+M/m_{\rm e})$, then  $\langle 24\,^3P_1| B_{X}(M)^\prime |24\,^3P_1\rangle = -1.854\,878\,538\,847(7)$.  This agrees with the value $-1.854\,878\,538\,8401(2)$ calculated independently by Tang et al.\ \cite{Tang2025}
 by the correlated B-spline method \cite{Chi2025,Tang2025}.

\begin{table*}[!tb]
\caption{Variational upper bounds (first entry) and extrapolations (second entry) for the nonrelativistic energies of the
$n\,^1P$ and $n\,^3P$ states of helium, assuming infinite nuclear mass. $N$ is the number of terms in the basis set. Units are atomic units.}
\label{bounds}
\begin{ruledtabular} \begin{tabular}{rrll}
\multicolumn{1}{c}{$n$} & \multicolumn{1}{c}{$N$}  &\multicolumn{1}{c}{$E(n\,^1P$)} & \multicolumn{1}{c}{$E(n\,^3P$)}\\
\hline
4 & 2775 &$  -2.03106\,96504\,50240\,71468\,1       $&$ -2.03232\,43542\,96630\,33188\,0       $\\  
  &      &$  -2.03106\,96504\,50240\,71474\,2(12)   $&$ -2.03232\,43542\,96630\,33195\,6(9)    $\\  
5 & 2775 &$  -2.01990\,59899\,00846\,44880\,6       $&$ -2.02055\,11872\,56267\,78819\,2       $\\  
  &      &$  -2.01990\,59899\,00846\,44884\,8(7)    $&$ -2.02055\,11872\,56267\,78824\,6(11)   $\\  
6 & 2622 &$  -2.01383\,39796\,71740\,10233\,5       $&$ -2.01420\,79587\,73750\,59118\,0       $\\  
  &      &$  -2.01383\,39796\,71740\,10238\,6(9)    $&$ -2.01420\,79587\,73750\,59121\,9(6)    $\\  
7 & 2949 &$  -2.01016\,93145\,29388\,98487\,7       $&$ -2.01040\,49600\,07971\,43124\,2       $\\  
  &      &$  -2.01016\,93145\,29388\,98489\,8(3)    $&$ -2.01040\,49600\,07971\,43126\,2(5)    $\\  
8 & 2839 &$  -2.00778\,91271\,33235\,89513\,8       $&$ -2.00794\,70137\,71161\,50524\,9       $\\  
  &      &$  -2.00778\,91271\,33235\,89515\,6(3)    $&$ -2.00794\,70137\,71161\,50526\,2(5)    $\\  
9 & 2864 &$  -2.00615\,63846\,52853\,81832\,7       $&$ -2.00626\,72673\,66409\,03255\,4       $\\  
  &      &$  -2.00615\,63846\,52853\,81834\,6(5)    $&$ -2.00626\,72673\,66409\,03258\,0(14)   $\\  
10& 5303 &$  -2.00498\,79838\,02218\,23945\,8138    $&$ -2.00506\,88054\,97707\,31644\,539     $\\  
  &      &$  -2.00498\,79838\,02218\,23945\,8183(24)$&$ -2.00506\,88054\,97707\,31644\,544(3)  $\\  
11& 5821 &$  -2.00412\,31919\,22332\,65252\,893     $&$ -2.00418\,39031\,99590\,64237\,6319    $\\  
  &      &$  -2.00412\,31919\,22332\,65252\,887(4)  $&$ -2.00418\,39031\,99590\,64237\,6334(1) $\\  
12& 4868 &$  -2.00346\,52527\,04885\,79826\,025     $&$ -2.00351\,20065\,35142\,74556\,011     $\\  
  &      &$  -2.00346\,52527\,04885\,79826\,035(5)  $&$ -2.00351\,20065\,35142\,74556\,021(18) $\\  
13& 5588 &$  -2.00295\,30939\,58149\,78487\,052     $&$ -2.00298\,98597\,64908\,81632\,212(11) $\\  
  &      &$  -2.00295\,30939\,58149\,78487\,063(6)  $&$ -2.00298\,98597\,64908\,81632\,221(5)  $\\  
14& 5604 &$  -2.00254\,66253\,70190\,96801\,19097   $&$ -2.00257\,60564\,26625\,76904\,83      $\\  
  &      &$  -2.00254\,66253\,70190\,96801\,19159(29$&$ -2.00257\,60564\,26625\,76904\,86(3)   $\\  
15& 6160 &$  -2.00221\,86471\,04088\,30158\,821     $&$ -2.00224\,25712\,22150\,32643\,875     $\\  
  &      &$  -2.00221\,86471\,04088\,30158\,826(6)  $&$ -2.00224\,25712\,22150\,32643\,884(5)  $\\  
16& 6160 &$  -2.00195\,01779\,73979\,34078\,79      $&$ -2.00196\,98874\,03296\,96697\,712     $\\  
  &      &$  -2.00195\,01779\,73979\,34078\,87(95)  $&$ -2.00196\,98874\,03296\,96697\,718(7)  $\\  
17& 5518 &$  -2.00172\,76459\,99910\,51694\,4161    $&$ -2.00174\,40751\,91087\,35600\,490     $\\  
  &      &$  -2.00172\,76459\,99910\,51694\,4179(27)$&$ -2.00174\,40751\,91087\,35600\,498(14) $\\  
18& 5518 &$  -2.00154\,11387\,60913\,62470\,8305    $&$ -2.00155\,49769\,40232\,72665\,171     $\\  
  &      &$  -2.00154\,11387\,60913\,62470\,8318(8) $&$ -2.00155\,49769\,40232\,72665\,182(4)  $\\  
19& 5870 &$  -2.00138\,32801\,02341\,60978\,0542    $&$ -2.00139\,50446\,04295\,12498\,77      $\\  
  &      &$  -2.00138\,32801\,02341\,60978\,0599(16)$&$ -2.00139\,50446\,04295\,12498\,76(3)   $\\
20& 5624 &$  -2.00124\,84894\,50687\,61310\,873     $&$ -2.00125\,85746\,92820\,42380\,855     $\\  
  &      &$  -2.00124\,84894\,50687\,61310\,887(11) $&$ -2.00125\,85746\,92820\,42380\,893(9)  $\\  
21& 6744 &$  -2.00113\,24817\,33017\,52951\,735     $&$ -2.00114\,11926\,56477\,67869\,1362    $\\  
  &      &$  -2.00113\,24817\,33017\,52951\,738(2)  $&$ -2.00114\,11926\,56477\,67869\,1390(23)$\\  
22& 6744 &$  -2.00103\,19225\,52162\,28543\,55462   $&$ -2.00103\,94979\,12816\,70120\,9912    $\\  
  &      &$  -2.00103\,19225\,52162\,28543\,55516(21$&$ -2.00103\,94979\,12816\,70120\,9925(7) $\\  
23& 6180 &$  -2.00094\,41858\,77955\,80644\,3933    $&$ -2.00095\,08147\,60962\,14631\,450     $\\  
  &      &$  -2.00094\,41858\,77955\,80644\,3957(7) $&$ -2.00095\,08147\,60962\,14631\,469(6)  $\\  
24& 6744 &$  -2.00086\,71808\,46170\,11128\,219     $&$ -2.00087\,30145\,66616\,65939\,225     $\\  
  &      &$  -2.00086\,71808\,46170\,11128\,223(6)  $&$ -2.00087\,30145\,66616\,65939\,240(9)  $\\  
25& 6744 &$  -2.00079\,92260\,24103\,06304\,5478    $&$ -2.00080\,43868\,29929\,07060\,8333    $\\  
  &      &$  -2.00079\,92260\,24103\,06304\,5555(6) $&$ -2.00080\,43868\,29929\,07060\,8458(13)$\\  
26& 6744 &$  -2.00073\,89568\,37741\,71921\,753     $&$ -2.00074\,35443\,60330\,29581\,243     $\\  
  &      &$  -2.00073\,89568\,37741\,71921\,765(4)  $&$ -2.00074\,35443\,60330\,29581\,281(17) $\\  
27& 6744 &$  -2.00068\,52565\,28882\,40212\,1288    $&$ -2.00068\,93526\,23570\,97655\,851     $\\  
  &      &$  -2.00068\,52565\,28882\,40212\,1453(18)$&$ -2.00068\,93526\,23570\,97655\,876(3)  $\\  
28& 6744 &$  -2.00063\,72040\,47170\,83714\,2907    $&$ -2.00064\,08764\,67041\,02409\,2930    $\\  
  &      &$  -2.00063\,72040\,47170\,83714\,3041(29)$&$ -2.00064\,08764\,67041\,02409\,3026(20)$\\  
29& 6744 &$  -2.00059\,40342\,90981\,60178\,88      $&$ -2.00059\,73395\,04545\,63185\,36      $\\  
  &      &$  -2.00059\,40342\,90981\,60178\,91(1)   $&$ -2.00059\,73395\,04545\,63185\,45(2)   $\\  
30& 6744 &$  -2.00055\,51074\,62372\,97425\,66      $&$ -2.00055\,80928\,35757\,97528\,28      $\\  
  &      &$  -2.00055\,51074\,62372\,97425\,91(24)  $&$ -2.00055\,80928\,35757\,97528\,32(5)   $\\  
31& 7948 &$  -2.00051\,98852\,24314\,85698\,88      $&$ -2.00052\,25907\,26604\,94616\,82      $\\  
  &      &$  -2.00051\,98852\,24314\,85698\,89(1)   $&$ -2.00052\,25907\,26604\,94616\,86(4)   $\\  
32& 7316 &$  -2.00048\,79119\,87756\,79932\,35      $&$ -2.00049\,03715\,34947\,75603\,30      $\\  
  &      &$  -2.00048\,79119\,87756\,79932\,41(1)   $&$ -2.00049\,03715\,34947\,75603\,39(3)   $\\  
33& 7948 &$  -2.00045\,88001\,04867\,46785\,88      $&$ -2.00046\,10426\,27369\,81858\,51      $\\  
  &      &$  -2.00045\,88001\,04867\,46785\,90(2)   $&$ -2.00046\,10426\,27369\,81858\,54(4)   $\\  
34& 7948 &$  -2.00043\,22180\,63626\,95660\,94      $&$ -2.00043\,42683\,60440\,67655\,20      $\\  
  &      &$  -2.00043\,22180\,63626\,95660\,98(4)   $&$ -2.00043\,42683\,60440\,67655\,23(3)   $\\  
35& 7948 &$  -2.00040\,78810\,08092\,82096\,5       $&$ -2.00040\,97604\,35014\,70404\,3       $\\  
  &      &$  -2.00040\,78810\,08092\,82096\,8(2)    $&$ -2.00040\,97604\,35014\,70404\,5(2)    $\\  
\end{tabular}
\end{ruledtabular}\\
\end{table*}

\begin{table*}[!tb]
\caption{Matrix elements of the spin-dependent operators in the Breit interaction for the 
$n\,^1P$ and $n\,^3P$ states of helium, calculated for the case $J=1$ \footnote{excluding the anomalous magnetic moment correction.}. 
For each pair, the first entry is $\langle n\,^3P_1|B_{X,\infty}^\prime| n\,^3P_1\rangle$ and
the second is the mass polarization correction coefficient $\langle n\,^3P_1|B_{X,M}^\prime| n\,^3P_1\rangle$, in units of $10^{-5}\alpha^2$ a.u.}
\label{matrix}
\begin{ruledtabular}
\begin{tabular}{rllll}
  $n$&\multicolumn{1}{c}{$\langle n\,^3P_1|B_{\rm so}^\prime|n\,^3P_|\rangle$}&
      \multicolumn{1}{c}{$\langle n\,^3P_1|B_{\rm soo}^\prime| n\,^3P_|\rangle$}&
      \multicolumn{1}{c}{$\langle n\,^3P_1|B_{\rm ss}^\prime| n\,^3P_|\rangle$}&
      \multicolumn{1}{c}{$\langle n\,^3P_1|\Delta_{3Z}^\prime| n\,^3P_1\rangle$}\\
\hline  
  24 &$  -1.854352474671(9)    $&   2.674998240246(7)   &$   -1.0876431688957(15)  $&   5.4271884889(18)  \\
     &$  -3.83779574(5)        $&   4.92099467099(2)    &$   -1.426091640(5)       $&   7.233732(9)       \\
  25 &$  -1.640050608267(9)    $&   2.365817143049(9)   &$   -0.9618981852881(32)  $&   4.7999883364(5)   \\
     &$  -3.39005620(5)        $&   4.34621403(5)       &$   -1.258871388(19)      $&   6.3850253(24)     \\
  26 &$  -1.457538177043(19)   $&   2.102506767458(24)  &$   -0.854814797824(6)    $&   4.26582611285(20) \\
     &$  -3.00935205(10)       $&   3.85758656(15)      &$   -1.116808321(39)      $&   5.6641036(13)     \\
  27 &$  -1.3011326394968(16)  $&   1.876866053129(9)   &$   -0.7630553854971(8)   $&   3.8080708978(9)   \\
     &$  -2.683591552(7)       $&   3.43955493(5)       &$   -0.9953436324(24)     $&   5.047800(6)       \\
  28 &$  -1.166328089046(7)    $&   1.682392385106(22)  &$   -0.683973821695(8)    $&   3.4135353602(4)   \\
     &$  -2.40320799(5)        $&   3.07981395(10)      &$   -0.89087475(6)        $&   4.5177234(29)     \\
  29 &$  -1.049520545597(11)   $&   1.5138849770276(27) &$   -0.6154539294841(11)  $&   3.07167179(8)     \\
     &$  -2.160566619(11)      $&   2.7685477611(34)    &$   -0.8005302417(12)     $&   4.0580(6)         \\
  30 &$  -0.947804167335(10)   $&   1.36715066900(7)    &$   -0.555789670118(24)   $&   2.773975696(11)   \\
     &$  -1.94952534(6)        $&   2.4978561(5)        &$   -0.72200070(18)       $&   3.66098(8)        \\
  31 &$  -0.8588184319796(30)  $&   1.2387833248071(37) &$   -0.5035953168823(11)  $&   2.51353855043(9)  \\
     &$  -1.765101303(13)      $&   2.2613382989(28)    &$   -0.6534165562(24)     $&   3.31315810(27)    \\
  32 &$  -0.7806323832762(37)  $&   1.125996743532(7)   &$   -0.457737428053(7)    $&   2.28470903(5)     \\
     &$  -1.6032268020(27)     $&   2.0537662962(7)     &$   -0.59325170914(36)    $&   3.00724(34)       \\
  33 &$  -0.711656127544(10)   $&   1.0264969147322(15) &$   -0.4172828436802(5)   $&   2.08283398(11)    \\
     &$  -1.460557904(28)      $&   1.87084326755(8)    &$   -0.54025268262(11)    $&   2.7381(8)         \\
  34 &$  -0.65057257124(16)    $&   0.9383834934(15)    &$   -0.3814585816(8)      $&   1.9040587(4)      \\
     &$  -1.3343283(12)        $&   1.709018(11)        &$   -0.493384(6)          $&   2.5052(29)        \\
  35 &$  -0.5962843365(25)     $&   0.8600731859(6)     &$   -0.34962065507060(6)  $&   1.745171(4)       \\
     &$  -1.222212(18)         $&   1.565333(4)         &$   -0.4517884294(5)      $&   2.263(29)         \\
\end{tabular}
\end{ruledtabular}
\end{table*}

$^*$Email address: gdrake@uwindsor.ca


\begin{references}
\bibitem{data}
Further tables of matrix elements for all $n\,P$ states of Helium
with $2\le n \le 35$ are available at DOI: 10.5061/dryad.cjsxksnkv.
\end{references}
\end{document}